\documentclass[conference]{IEEEtran}
\usepackage{cite}
\usepackage{multirow}
\usepackage{amsmath,amssymb,amsfonts}
\usepackage{algorithmic}
\usepackage{graphicx}
\usepackage{caption}
\usepackage{subcaption}
\usepackage{textcomp}
\usepackage{xcolor}
\usepackage[hyphens]{url}
\usepackage{fancyhdr}
\usepackage{array}      
\usepackage{tabularx}   
\newcolumntype{C}{>{\centering\arraybackslash}X} 
\usepackage[bookmarks=true,breaklinks=true,letterpaper=true,colorlinks,citecolor=black,linkcolor=black,urlcolor=black,hyperfootnotes=false]{hyperref}

\begin{document}
\IEEEoverridecommandlockouts
\title{Dissecting and Re-architecting 3D NAND Flash PIM Arrays for Efficient Single-Batch Token Generation in LLMs}
\author{
    Yongjoo Jang$^{1}$, Sangwoo Hwang$^{1}$, Hojin Lee$^{1}$, Sangwoo Jung$^{1}$, Donghun Lee$^{1}$, Wonbo Shim$^{2}$, Jaeha Kung$^{1, \dagger \thanks{$\dagger$ J. Kung is the corresponding author (e-mail: jhkung@korea.ac.kr).}}$\\
    {\small \textit{$^1$Korea University, Seoul, South Korea, 
    $^2$Seoul National University of Science and Technology, Seoul, South Korea
    }} \\
    {\small \textit{\{jyjoo, nemesis0523, hojin5344, swjung1, dhleeids, jhkung\}@korea.ac.kr, wbshim@seoultech.ac.kr}}
}
\maketitle
\pagestyle{plain}

\begin{abstract}
The advancement of large language models has led to models with billions of parameters, significantly increasing memory and compute demands. 
Serving such models on conventional hardware is challenging due to limited DRAM capacity and high GPU costs. 
Thus, in this work, we propose offloading the single-batch token generation to a 3D NAND flash processing-in-memory (PIM) device, leveraging its high storage density to overcome the DRAM capacity wall.
We explore 3D NAND flash configurations and present a re-architected PIM array with an H-tree network for optimal latency and cell density. 
Along with the well-chosen PIM array size, we develop operation tiling and mapping methods for LLM layers, achieving a 2.4$\times$ speedup over four RTX4090 with vLLM and comparable performance to four A100 with only 4.9\% latency overhead. 
Our detailed area analysis reveals that the proposed 3D NAND flash PIM architecture can be integrated within a 4.98mm$^2$ die area under the memory array, without extra area overhead.
\end{abstract}


\section{Introduction}\label{sec:intro}
\begin{figure}[t]
\begin{subfigure}{0.48\linewidth}
\centerline{\includegraphics[width=\columnwidth]{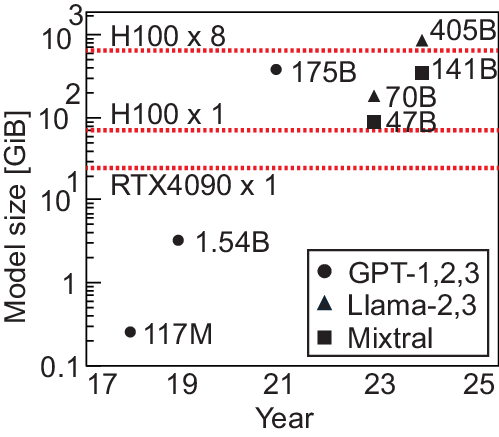}}
\caption{LLM memory requirement}
\label{fig:llm_memory}
\end{subfigure}
\hspace{0.01\linewidth}
\begin{subfigure}{0.48\linewidth}
\centerline{\includegraphics[width=\columnwidth]{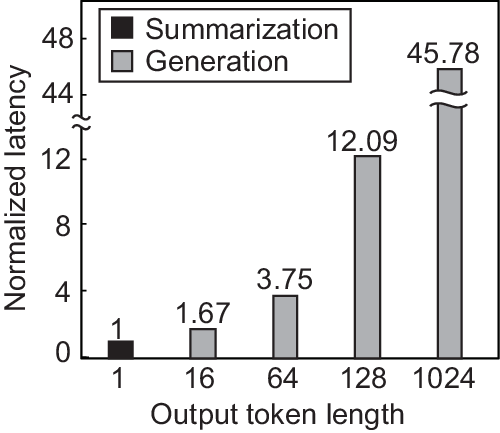}}
\caption{Token generation latency}
\label{fig:llm_gen_latency}
\end{subfigure}
\caption{Challenges in LLM token generation: (a) substantial memory requirements and (b) higher token generation latency than summarization (OPT-30B on 4$\times$RTX4090).}\vspace{-2mm}
\end{figure}

The proposal of transformer architecture~\cite{transformer} in 2017 has ignited the development of various large language models (LLMs)~\cite{gpt3,mixtral}. 
As emergent abilities of LLMs have been explored, the size of LLMs has increased significantly to learn more complex language syntax and semantics.
Thus, running LLMs on commodity hardware poses significant challenges due to their high computational and memory requirements. 
Specifically, these models often contain billions of parameters, demanding substantial GPU memory (Fig.~\ref{fig:llm_memory}).
For instance, Mixtral consists of eight experts having 47B parameters, which is built upon Mistral-7B~\cite{mixtral}.
The number of parameters $N$ translates into 2-Byte$\times N$ memory capacity in \texttt{FP16} or \texttt{BF16}, requiring 94GiB for serving a 47B model, which exceeds NVIDIA's H100 DRAM capacity, i.e., 80GiB.
Considering the high cost of an H100 card, 
serving a single model with two H100 cards puts a lot of pressure on LLM service providers.
Even worse, GPT-3.5 uses 175B parameters~\cite{gpt3}, requiring 350GiB as its storage, which translates to five H100 GPUs.

To overcome the capacity limitation of DRAM in servicing a large-scale LLM, one may think of utilizing a 3D NAND flash as a direct storage of model parameters.
By stacking several hundreds of NAND flash layers~\cite{nand_isscc25}, the cell density of NAND flash is significantly higher than that of DRAM.
However, moving data back and forth from the storage to the computing fabric is limited by its poor PCIe bandwidth.
In LLM inference, the data bandwidth directly impacts the latency since a massive number of model parameters needs to be fetched from the memory to predict the next token.
In this work, we try to leverage processing-in-memory (PIM) technology using 3D NAND flash, i.e., processing-in-flash, for high-performance and cost-effective token generation in LLMs with a conventional PCIe-based storage system (\textit{scalable}).

\begin{figure*}[t]
\begin{subfigure}{0.49\linewidth}
\centerline{\includegraphics[width=0.8\textwidth]{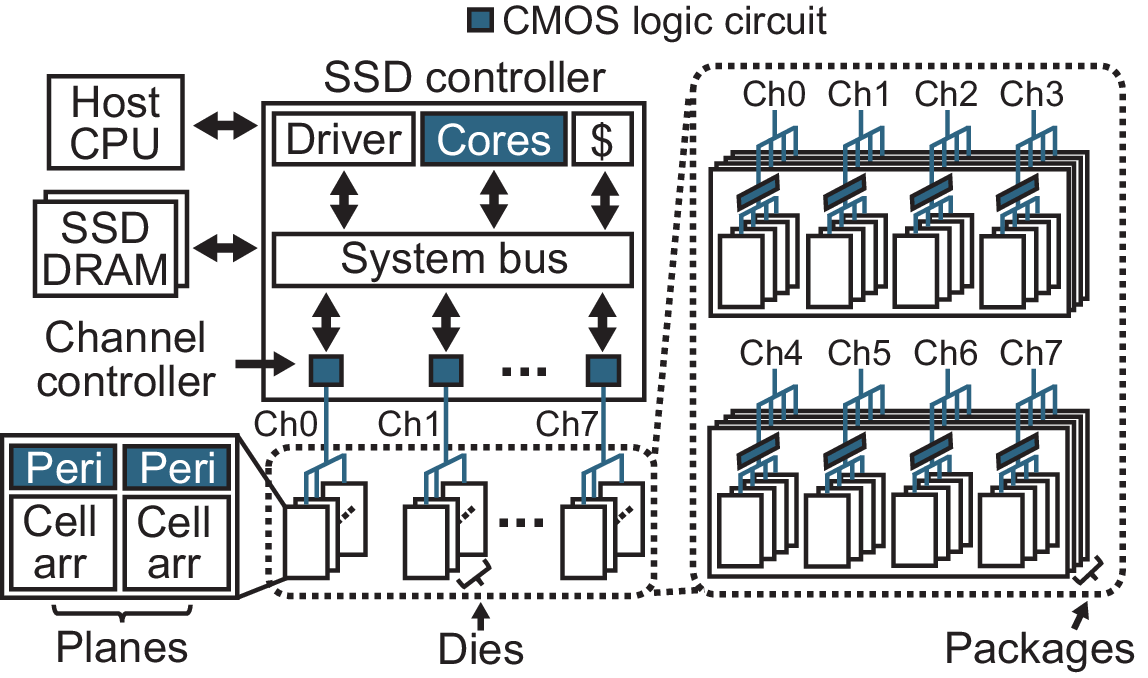}
}
\caption{NAND flash architecture}
\label{fig:nand_hierarchy}
\end{subfigure}
\begin{subfigure}{0.49\linewidth}
\centerline{\includegraphics[width=0.8\textwidth]{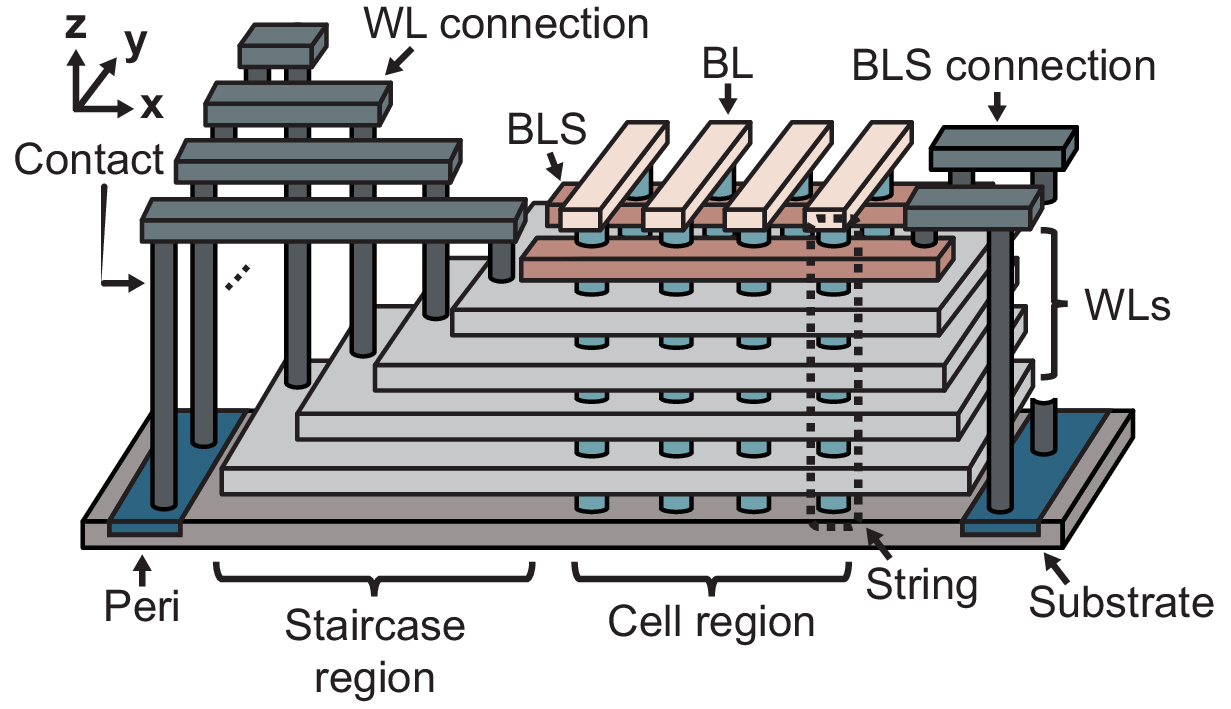}
}
\caption{Bird-eye view of a 3D NAND flash plane}
\label{fig:nand_plane_birdeye}
\end{subfigure}
\caption{(a) A hierarchical NAND flash architecture from memory cell arrays to an SSD controller. (b) A plane consists of a 3D memory cell array and peripheral circuits. To activate one of 3D-stacked wordlines (WLs), a staircase region is used to allow multi-layer WL connections. A bitline (BL) and a bitline select (BLS) intersects at a string.}\vspace{-2mm}
\end{figure*}

For generative tasks, there are two important steps involved: summarization and token generation.
Fig.~\ref{fig:llm_gen_latency} demonstrates that the latency of generating 1K output tokens consumes 46$\times$ higher latency than summarizing 1K input tokens with OPT-30B running on four RTX4090.
This is due to the lower arithmetic intensity of the token generation compared to the summarization stage.
The low arithmetic intensity makes the generation stage bottlenecked by the memory bandwidth.
To mitigate this, batching multiple user requests can be done, which incurs KV caching overhead~\cite{vllm}.
Instead of pressing GPUs to handle multi-batch summarization and generation, we propose to \textit{assign single-batch generation task to a flash PIM device so that GPUs are released for other summarization requests}.
This architectural choice only requires the initial key-value (KV) cache of input tokens to be moved from GPU DRAM to NAND flash through the PCIe interface.

There are several studies that utilize 3D NAND flash PIM as a store-and-compute device for deep learning~\cite{flash_pim_ucsb, jssc_flash_pim,3d-fpim}.
In~\cite{flash_pim_ucsb}, the authors have proposed a time-domain matrix-vector multiplication (MVM) scheme using 3D flash.
Activations are encoded as voltage pulses, with pulse widths proportional to their values.
Since it uses time-domain computation, however, the compute latency increases to guarantee high precision.
In~\cite{jssc_flash_pim}, embedded 3D flash PIM has been fabricated using standard logic processes to demonstrate its ability to run convolutional neural networks (CNNs).
However, it demonstrates a PIM functionality without optimizing the NAND flash array size for its optimal performance in terms of latency, energy, and area efficiency.
3D-FPIM~\cite{3d-fpim} proposes several architectural features that improve the energy efficiency of 3D flash PIM, i.e., quantization-aware ADC, multi-stack MVM, and wordline reuse (but only applied to CNNs; the dynamic range of partial sums in CNNs is smaller than that of LLMs).
Similar to~\cite{jssc_flash_pim}, 3D-FPIM fixes the array size to 128$\times$128 without exploring the design space of 3D NAND flash PIM.

Thus, in this work, we dissect and re-architect the 3D NAND flash PIM arrays to maximize the performance of single-batch token generation with minimal memory density loss. 
We first find the optimal size of a 3D flash PIM array, then design a PIM-enabled NAND flash architecture with an H-tree network.
This work makes the following key contributions:
\begin{itemize}
\item We select \textit{the optimal 3D NAND flash array size} for low-latency PIM operations, i.e., 2$\mu$s, while keeping high cell density. With a new plane configuration, the \textit{H-tree bus architecture is utilized} to fully parallelize PIM arrays.
\item We classify MVM operations into two categories: static- and dynamic-MVMs. Then, \textit{the operation mapping and dataflow for these two MVMs are explored} in detail.
\item We \textit{present a QLC-SLC hybrid architecture} so that KV caching and dMVM are efficiently supported on a flash.
\end{itemize}

\section{Preliminaries}\label{sec:prelim}
\subsection{3D NAND Flash Architecture}\label{sec:flash_arch}
A typical NAND flash architecture has a hierarchy, as shown in Fig.~\ref{fig:nand_hierarchy}.
The example in Fig.~\ref{fig:nand_hierarchy} depicts a NAND flash with 8 channels, 4 ways (i.e., packages) per channel, 4 dies per way, and 2 planes per die (256 planes in total). 
The NAND flash channels can be accessed in parallel to improve read/write performance.
In addition, there are logic circuits that support the required operations at each channel, way, and plane.
At the lowest level of the hierarchy, there are planes with their own peripheral circuits.
Each plane has a 3D structure which consists of multiple strings, each with vertically stacked memory cells in a cylindrical shape (Fig.~\ref{fig:nand_plane_birdeye}).
Strings in the y-direction are connected via bitline (BL) at the top, while those in x-direction are connected via bitline select (BLS).
A page (e.g., 4KB) can be accessed by activating a wordline (WL) and a BLS by using dedicated peripheral circuits on the substrate.
To allow the activation of a single WL at a time, a staircase region is used.
Therefore, if the cell region, which is the actual memory array, becomes smaller, the memory density reduces due to the staircase overhead.

\begin{figure}[t]
\begin{subfigure}{0.48\linewidth}
\centerline{\scalebox{1}{\includegraphics[width=\textwidth]{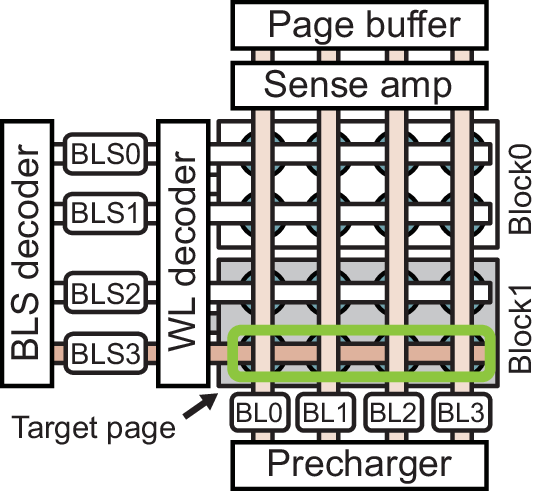}}}
\caption{Top view of a plane}
\label{fig:nand_page_read_L}
\end{subfigure}
\hspace{0.02\linewidth}
\begin{subfigure}{0.48\linewidth}
\centerline{\scalebox{1}{\includegraphics[width=\textwidth]{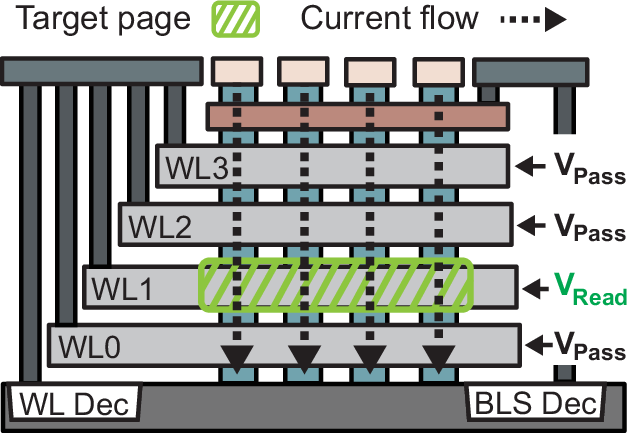}}}
\caption{Side view of a plane}
\label{fig:nand_page_read_R}
\end{subfigure}
\caption{(a) A top view and (b) a side view of a 3D NAND flash plane when a page read is performed.}\vspace{-2mm}
\label{fig:nand_read}
\end{figure}

Fig.~\ref{fig:nand_read} shows how the read operation is performed in a typical 3D NAND flash plane.
First, a WL decoder selects and drives a single block at a specific WL, i.e., `Block 1' in Fig.~\ref{fig:nand_page_read_L}, on which the target page is located. 
Then, the target page is accessed by driving the corresponding BLS (i.e., `BLS3') by using a BLS decoder.
The accessed data are transferred via BLs and decoded in the sense amplifier, which are finally latched in the page buffer.
To activate the target WL where the target page is located, i.e., `WL1' in Fig.~\ref{fig:nand_page_read_R}, the WL decoder drives the target with a read voltage ($V$\textsubscript{Read}) while the other WLs are driven by a pass voltage ($V$\textsubscript{Pass}).
The activated WL is highlighted in green in Fig.~\ref{fig:nand_page_read_R}.
Depending on the stored data, e.g., 0 or 1 for a single-level cell (SLC), current may flow from BL at the top of the string to the readout circuit on the substrate.
Thus, the page read latency of a NAND flash with SLCs can be expressed as:
\begin{equation}\label{eq:page_read}
T_{\text{read}} = t_{\text{decWL}} + \max(t_{\text{decBLS}}, t_{\text{pre}}) + t_{\text{sense}} + t_{\text{dis}},
\end{equation}
where $t_{\text{decWL}}$ and $t_{\text{decBLS}}$ are latencies to activate a WL/BLS via WL and BLS decoders, respectively.
The $t_{\text{pre}}$, $t_{\text{sense}}$, $t_{\text{dis}}$ are latencies for precharging BLs, sensing and latching a page, and discharging BLs for next operations, respectively.


\subsection{Operation of 3D NAND Flash PIM}\label{sec:flash_pim}
\begin{figure}[t]
\begin{subfigure}{0.50\linewidth}
\centerline{\scalebox{1}{\includegraphics[width=\columnwidth]{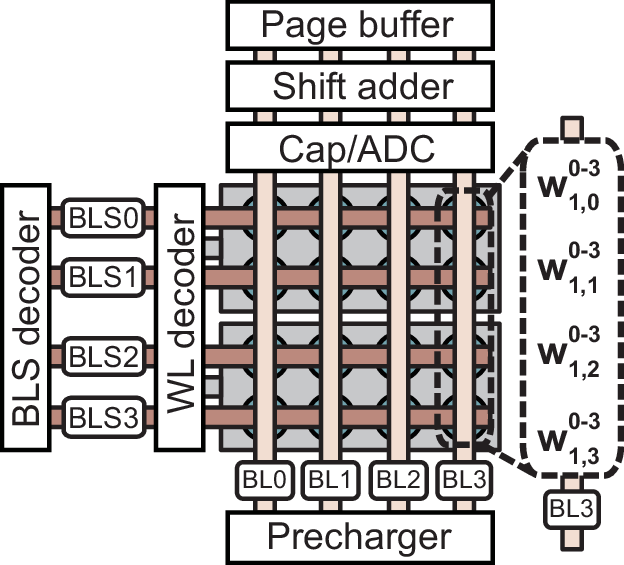}}}
\caption{PIM-enabled NAND plane}
\label{fig:nand_pim_plane_L}
\end{subfigure}
\hspace{0.02\linewidth}
\begin{subfigure}{0.46\linewidth}
\centerline{\scalebox{1.0}{\includegraphics[width=\columnwidth]{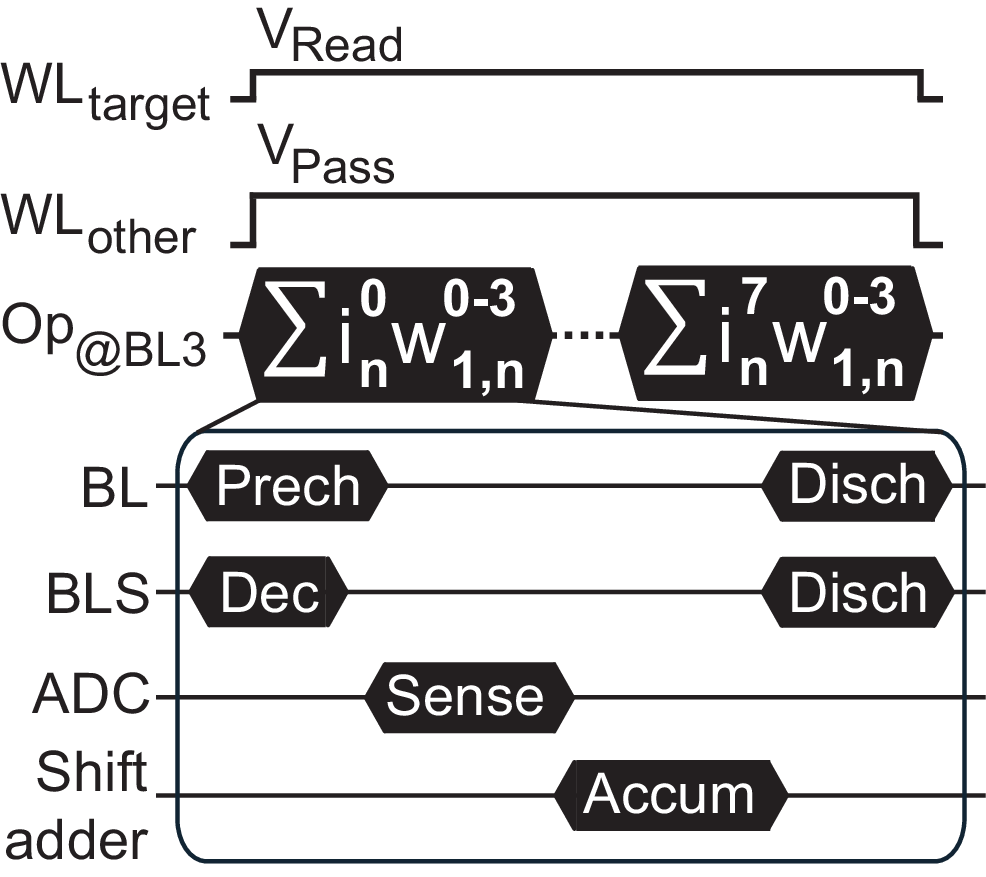}}}
\caption{Timing diagram of PIM}
\label{fig:nand_pim_plane_R}
\end{subfigure}
\caption{Simple example of 3D NAND flash PIM operation.}\vspace{-2mm}
\label{fig:nand_pim}
\end{figure}

The 3D NAND flash can be used as a processing-in-memory unit by placing additional peripheral circuits and simultaneously activating multiple blocks (Fig.~\ref{fig:nand_pim_plane_L}).
In the PIM mode, a dot product is performed by flowing current through the BL.
The dot product can be defined as:
\begin{equation}\label{eq:dot_product}
o_k = \sum_{b}\sum_{n} i_n^b \cdot \mathbf{w}_{k,n}=\sum_{b}i_0^b\cdot \mathbf{w}_{k,0} + ...+i_{N-1}^b\cdot \mathbf{w}_{k,N-1},
\end{equation}
where $o_k$ is the $k$\textsuperscript{th} output, $i_n^b$ is the $b$\textsuperscript{th}-bit of $n$\textsuperscript{th} input, and $\mathbf{w}_{k,n}$ is the multi-bit weight connecting $i_n$ and $o_k$.
A 4-bit $\mathbf{w}_{k,n}$ can be stored in a single quad-level cell (QLC).
To enable PIM, the WL decoder drives multiple blocks in the target WL instead of one with $V_\text{Read}$, while the other blocks/WLs are driven with $V_\text{Pass}$.
Due to reliability, the number of simultaneously activated blocks in QLC NAND flash PIM is limited to 256 cells being accumulated through a single BL~\cite{3d-fpim}.
Then, the BLS decoder drives BLSs depending on the input signal $i_n^b$ (0 or 1).
Note that a multi-bit input is handled in a time-sequential manner.
In Fig.~\ref{fig:nand_pim_plane_L}, we assume QLC cells with 8-bit inputs and weights.
Therefore, four 8-bit weights are stored across two BLs, e.g., BL2 ($\mathbf{w}_{1,n}^{4-7}$) and BL3 ($\mathbf{w}_{1,n}^{0-3}$) for $o_1$.
Then, $\sum_n i_n^0 \cdot \mathbf{w}_{1,n}^{0-3}$ is computed through BL3 at the first clock cycle (Fig.~\ref{fig:nand_pim_plane_R}).
The operation at BL3 is broken down into multiple steps: 
(i) WL decoding, (ii) BL precharge/BLS decoding, (iii) ADC sensing, (iv) accumulation, and (v) BL/BLS discharge.
The digitized results by ADC from BL2 and BL3 are accumulated together at the shift adder to obtain the final $o_1$.
Thus, the latency of 3D flash PIM can be defined as:
\begin{equation}
\begin{aligned}
T_{\text{PIM}} = t_{\text{decWL}}
  &+ \bigl(\max(t_{\text{decBLS}}, t_{\text{pre}}) + t_{\text{sense}} \\[2pt]
  &\qquad + t_{\text{accum}} + t_{\text{dis}}\bigr)\times B_{\text{input}},
\end{aligned}
\label{eq:pim_latency}
\end{equation}
where $t_{\text{accum}}$ is the result accumulation latency and $B_{\text{input}}$ is the bit-width of the input.

\section{Proposed 3D Flash PIM Architecture}\label{sec:proposed_arch}
\subsection{Challenges in Developing 3D Flash PIM}\label{sec:challenges}

\begin{figure}[t]
\centerline{{\includegraphics[width=0.8\columnwidth]{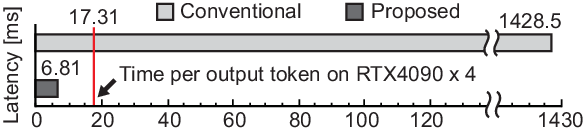}}}
\caption{Comparison of time per output token (i.e., TPOT) with OPT-30B between the conventional and the proposed 3D NAND PIM architecture.}\vspace{-2mm}
\label{fig:conventional_size_nand_pim}
\end{figure}

Directly using the same flash architecture for PIM has a major challenge in terms of latency.
Typically, a 3D NAND flash plane has 4 rows per block, 700-2,800 blocks per plane, and 64-128 stacks with 20-50$\mu$s read latency~\cite{isscc_18,isscc_19_pua}.
This substantial read latency introduces a significant delay in token generation when making a 3D flash PIM with the conventional plane size, as shown in Fig.~\ref{fig:conventional_size_nand_pim}.
It takes 1.4s to generate an output token with 8-bit quantized OPT-30B using the na\"ive implementation of 3D flash PIM.
By selecting the proper plane size (Section~\ref{sec:plane_arch}) and bus architecture within the die (Section~\ref{sec:bus_arch}), we can significantly improve the time required to generate an output token by 210$\times$ (i.e., 2.5$\times$ faster than four RTX4090 with vLLM).
To meet the low latency requirements of emerging applications, a new type of NAND flash, e.g., Z-NAND~\cite{zssd}, has been introduced that provides the read latency of 3$\mu$s with a reduced page size.
Reducing the page size shrinks the WL region, effectively reducing the load capacitance of WL, equivalently $t_{\text{decWL}}$ in Eq.~(\ref{eq:pim_latency}).
Other components that determine the plane size, i.e., the number of blocks and stacks, also affect the latency.
For instance, with more rows (= BLSs), the BL gets longer, which increases the $t_{\text{pre}}$ in Eq.~(\ref{eq:pim_latency}).

\begin{figure*}[t]
\centerline{{\includegraphics[width=0.95\linewidth]{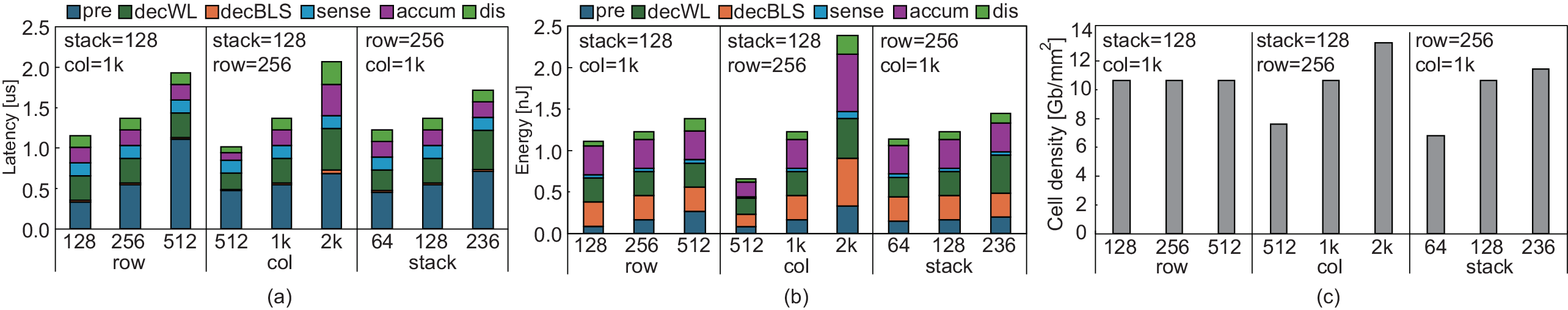}}}
\caption{(a) Latency (in $\mu$s), (b) energy consumption (in nJ), and (c) cell density (in Gb/mm\textsuperscript{2}) depending on the 3D NAND flash PIM configuration (i.e., number of BLSs, BLs, and stacks).}\vspace{-2mm}
\label{fig:latency_energy_celldensity_trend}
\end{figure*}


Unfortunately, there is a trade-off between the PIM latency and the cell density.
The cell density can be estimated by
\begin{equation}
  D_{\text{cell}} =
    \frac{N_{\text{col}}\,\times\,
          {N_\text{stack}}\,\times\,
          {B_\text{cell}}}
         {{L_\text{cell}}+
    {L_\text{staircase}}}\cdot\frac{N_\text{row}}{W},
\label{eq:cell_density}
\end{equation}
where $N_\text{col}$ and $N_\text{stack}$ are the number of BLs (page size), and stacks, $B_\text{cell}$ is the number of bits stored per cell, and $L_\text{cell}$ and $L_\text{staircase}$ is the length of the cell region and the staircase region, respectively.
The $W$ is the width of the plane, which is proportional to the number of rows $N_\text{row}$.
The $L_\text{staircase}$ increases as $N_\text{stack}$ gets higher, and this lowers the cell density if $L_\text{staircase}$ is comparable to $L_\text{cell}$. 
Thus, the $D_\text{cell}$ is high with the conventional plane size since $N_\text{col}$ is much higher than $N_\text{stack}$, which makes $L_\text{staircase}\ll L_\text{cell}$.
However, having a significantly large page size as usual increases the $T_\text{PIM}$ which conflicts with the design strategy of improving the cell density.
Thus, in the following section, we extensively explore the design space of 3D NAND flash PIM array size that balances the PIM latency and the cell density.

\subsection{3D NAND Plane Configuration}\label{sec:plane_arch}

A 3D NAND flash plane can be configured by `$N_\text{row}\times N_\text{col}\times N_\text{stack}$'.
These size parameters affect the lengths of BL, BLS, WL, and string, which change the resistance $R_x$ and the capacitance $C_x$ of them.
For instance, $R_\text{WL}$ and $C_\text{WL}$ are proportional to $L_\text{WL}=L_\text{cell}+L_\text{staircase}$.
Thus, with $R_x$ and $C_x$ calculated, we can estimate the latency of a 3D flash PIM for a given size configuration by
\begin{subequations}\label{eq:latency_block}
\begin{align}
t_{\text{pre}}
  &\approx h\bigl(\!R_{\text{s}} \!\times\!(N_{\text{col}}\!\cdot\!C_{\text{INV}}) \!\bigr)
  \!+\! h\bigl(\!R_{\text{BL}} \!\times\!(C_{\text{BL}}/2 \!+\! C_{\text{string}}) \!\bigr),\label{eq:t_pre} \\[2pt]
t_{\text{decBLS}}
  &\approx h\bigl(\!R_{\text{BLS}}\!\times\!C_{\text{BLS}}/2\!\bigr)\!,
     \label{eq:t_decBLS} \\[2pt]
t_{\text{decWL}}
  &\approx h\bigl(\!R_{\text{s}}\!\times\!(C_{\text{cell}}\!+\!C_{\text{stair}})\!\bigr),
     \label{eq:t_decWL}
\end{align}
\end{subequations}
where $h(\tau)\propto\tau^{\tiny1.5}$ is the Horowitz delay equation~\cite{horowitz}, where $\tau$ is the RC time constant (\textit{only dominant terms are shown}).
The precharge latency ($t_\text{pre}$) is mainly determined by the latency of turning on the $N_\text{col}$ precharge transistors through a switch transistor (1\textsuperscript{st} term) and the latency of precharging each BL (2\textsuperscript{nd} term).
The BLS decoding latency ($t_\text{decBLS}$) is dominated by $R_\text{BLS}\times C_\text{BLS}$ when $N_\text{col}$ is less than 16K.
The WL decoding latency ($t_\text{decWL}$) is dominated by switching on the pass transistor that drives the WL.
For $N_\text{stack}=128$, $C_\text{stair}$ is comparable to $C_\text{cell}$ with $N_\text{col}=512$.

We have modified the open-source 3D-FPIM simulator~\cite{3d-fpim} to incorporate 4:1 column multiplexers, 9-bit SAR-ADCs, and shift adders for accurate latency/power estimation.
By varying the $N_\text{row}$, $N_\text{col}$, and $N_\text{stack}$, we evaluated the latency, energy consumption, and cell density (Fig.~\ref{fig:latency_energy_celldensity_trend}).
The latency in Fig.~\ref{fig:latency_energy_celldensity_trend}a assumes that 128 BLSs are used to perform a dot product with QLCs, and both input and weight values are 8-bit (i.e., two neighboring flash cells are used to store an 8-bit weight).
One of the configuration parameters, e.g., $N_\text{row}$, is being changed while the remaining two are fixed ($N_\text{col}=1\text{K}$ and $N_\text{stack}=128$).
Obviously, as $N_\text{row}$, $N_\text{col}$, or $N_\text{stack}$ becomes larger, the PIM latency increases due to increased RC values in Eq.~(\ref{eq:latency_block}).
The precharge latency $t_\text{pre}$ sharply increases as $N_\text{row}$, i.e., the number of BLSs, increases since both $R_\text{BL}$ and $C_\text{BL}$ in Eq.~(\ref{eq:t_pre}) are proportional to $N_\text{row}$ (thus, $\tau_{BL}\propto N_\text{row}^2$).
Even though $t_\text{decBLS}$ has a similar impact when $N_\text{col}$ increases, it takes up a small portion of the total PIM latency since BLS is made of tungsten, having much lower $R$ and $C$ values than BL made of copper~\cite{book_3dflash}.
The WL decoding latency $t_\text{decWL}$ remains the same even with the increased $N_\text{row}$ since the number of simultaneously activated blocks is fixed for each dot product.
The $t_\text{decWL}$ has the sub-linear dependence on $N_\text{col}$ or $N_\text{stack}$ since each sizing parameter affects either $C_\text{cell}$ or $C_\text{stack}$.

As shown in Fig.~\ref{fig:latency_energy_celldensity_trend}b, the energy consumption of 3D flash PIM also increases as $N_\text{row}$, $N_\text{col}$, and $N_\text{stack}$ become larger.
The energy consumptions for some operations are defined as
\begin{subequations}\label{eq:energy_block}
\begin{align}
E_{\text{pre}}
  &\approx N_{\text{col}}\!\times\! V_{\text{pre}}^{2}\!\times\!\bigl(C_{\text{BL}}\!+\!C_{\text{string}}\!\times\!N_\text{row}^{*}\!\times\!(1\!-\!\alpha_{i})\!\bigr),
     \label{eq:E_pre} \\[2pt]
E_{\text{decBLS}}
  &\approx N_{\text{row}}^{*}\!\times\!V_\text{pass}^{2}\!\times\!C_{\text{BLS}},
     \label{eq:E_BLSDec} \\[2pt]
E_{\text{decWL}}
  &\approx V_\text{read}^{2}\!\times\!(C_{\text{cell}}\!+\!C_{\text{stair}})
    \!+\! V_\text{pass}^{2}\!\times\!(C_{\text{cell}}\!+\!C_{\text{stair}}),
     \label{eq:E_WLDec_SR} 
\end{align}
\end{subequations}
where $N_\text{row}^*$ is the number of simultaneously activated rows for each PIM operation and $\alpha_{i}$ is the sparsity of input bits at BLS. 
The $V_\text{pre}$, $V_\text{pass}$, and $V_\text{read}$ are precharge, pass, and read voltages, respectively.
According to Eq.~(\ref{eq:E_pre}), the energy consumption of BL precharge increases linearly with $N_\text{col}$ and $N_\text{row}$.
For our LLM benchmarks, the value of input bit sparsity is near 0.5, which makes $C_{\text{BL}}$ dominant over the $C_{\text{string}}$ term.
In contrast, the energy consumed by the BLS decoder is only proportional to $N_\text{col}$, which increases $C_\text{BLS}$.
The $N_\text{row}^*$ is fixed to 128 in Eq.~(\ref{eq:E_BLSDec}), making $E_\text{decBLS}$ irrelevant to $N_\text{row}$.
Similar to the PIM latency, the energy consumption of the WL decoder has a sub-linear dependence on both $N_\text{col}$ and $N_\text{stack}$.
The accumulation energy (\texttt{accum} in Fig.~\ref{fig:latency_energy_celldensity_trend}b) sharply increases with a higher $N_\text{col}$ as the controller needs to drive higher MUX loads.

As described above, reducing the size of a 3D NAND plane improves the compute speed and energy efficiency.
However, having a smaller plane leads to a reduced cell density according to Eq.~(\ref{eq:cell_density}).
Thus, we need to carefully set the plane configuration so as not to sacrifice the cell density too much while minimizing the latency.
Note that the cell density is irrespective of $N_\text{row}$ because $W$ is proportional to $N_\text{row}$ in Eq.~(\ref{eq:cell_density}).
The cell density is more sensitive to $N_\text{col}$ than to $N_\text{stack}$ because $L_\text{cell}$ is relatively smaller than $L_\text{staircase}$ for the simulated configurations (Fig.~\ref{fig:latency_energy_celldensity_trend}c).
If $N_\text{col}$ is much larger, e.g., 16K, the cell density will be more sensitive to $N_\text{stack}$ than $N_\text{col}$.
According to our detailed analysis, \textit{we select the plane size as} `$256\!\times\!2048\!\times\!128$', i.e., $N_\text{row}\!\times\!N_\text{col}\!\times\!N_\text{stack}$, for the maximum cell density while achieving $\sim$2$\mu$s PIM latency.

\subsection{Bus Architecture Within NAND Die}\label{sec:bus_arch}
\begin{figure}[t]
\begin{subfigure}{0.40\linewidth}
\centerline{\includegraphics[width=0.5\columnwidth]{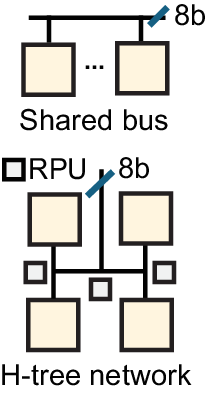}}
\caption{Shared bus vs. H-tree}
\label{fig:bus_structure_compare}
\end{subfigure}
\hspace{0.01\linewidth}
\begin{subfigure}{0.54\linewidth}
\centerline{{\includegraphics[width=0.9\columnwidth]{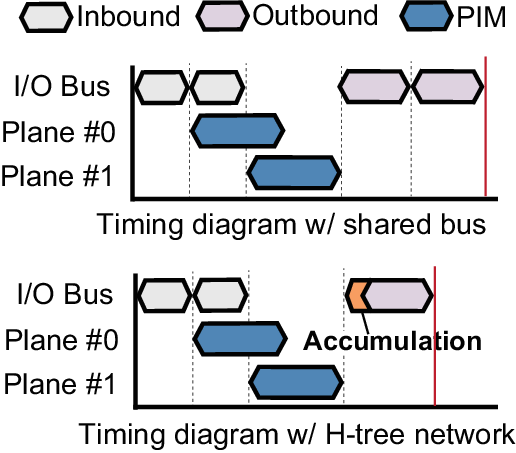}}}
\caption{Latency comparison}
\label{fig:bus_latency_compare_diagram}
\end{subfigure}
\caption{Latency comparison between PIM with a shared bus and PIM with an H-tree network.}\vspace{-2mm}
\end{figure}

\begin{figure}[t]
\centerline{{\includegraphics[width=0.90\columnwidth]{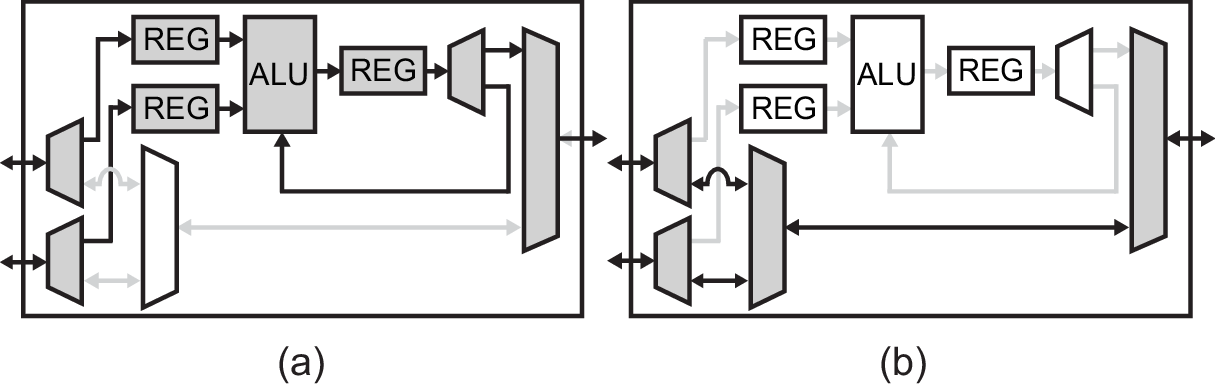}}}
\caption{Reconfigurable processing unit with two operation modes: (a) ALU mode and (b) stream mode.}\vspace{-2mm}
\label{fig:rpu}
\end{figure}

In the conventional flash die, planes are connected through a shared bus, as shown in Fig.~\ref{fig:bus_structure_compare}.
This saves the interconnect resources since only one plane is accessed at a time for regular read/write instructions.
For PIM operations, however, multiple planes can be activated at the same time, generating more output data.
With the bus speed of 1.6-2GB/s~\cite{networkedSSD}, PIM latency (e.g., 1-2$\mu$s in Fig.~\ref{fig:latency_energy_celldensity_trend}a) is much higher than I/O latency (i.e., 64ns for moving 128 8-bit data).
To reduce the latency gap between PIM and I/O instructions, we can pipeline PIM executions across multiple planes (Fig.~\ref{fig:bus_latency_compare_diagram}).
However, multiple outputs must travel outside the die for their accumulation, increasing I/O latency.
Instead, an H-tree bus topology can be used so that outputs are accumulated on the way to the output bus.
The accumulation is done using an ALU mode (Fig.~\ref{fig:rpu}a) of a reconfigurable processing unit (RPU), which takes outputs from two planes.
The stream mode of an RPU is used for regular read/write or program operations (Fig.~\ref{fig:rpu}b).
The overhead of realizing the H-tree network within a die will be discussed in Section~\ref{sec:exp_results}.


\begin{figure}[t]
\centerline{{\includegraphics[width=0.95\columnwidth]{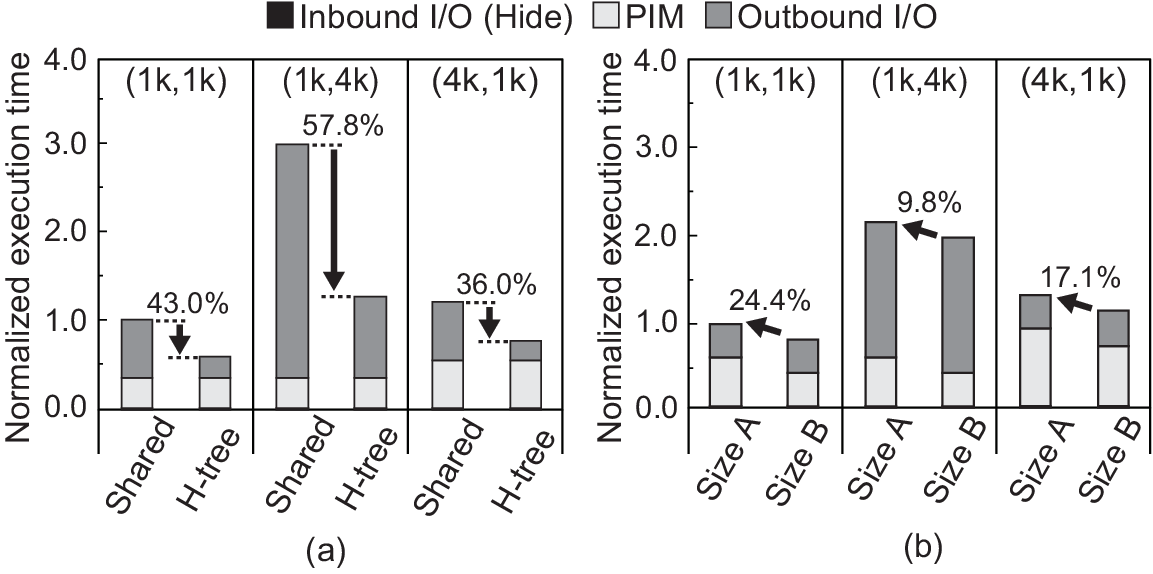}}}
\caption{The comparison of execution time (a) between shared-bus and H-tree with the same plane size, and (b) between two different plane sizes with H-tree bus network.}\vspace{-2mm}
\label{fig:htree_and_sharedbus_exeuction_time}
\end{figure}

To demonstrate the effectiveness of pipelined PIM execution with the H-tree network, three MVM cases are evaluated, i.e., $(1,M)\!\times\!(M,N)$, where $(M,N)$ is (1K,1K), (1K,4K) or (4K,1K).
For the evaluation, 64 planes are used for PIM, each with the size of $256\!\times\!2048\!\times\!128$ (\texttt{Size A}) which is selected in Section~\ref{sec:plane_arch}.
As shown in Fig.~\ref{fig:htree_and_sharedbus_exeuction_time}a, 46\% execution time reduction, on average, can be achieved by using the H-tree network.
We may reduce the execution time even further by using a smaller PIM, e.g., $256\!\times\!1024\!\times\!64$ (\texttt{Size B}), at the cost of the lower cell density (Fig.~\ref{fig:latency_energy_celldensity_trend}c).
By overlapping the PIM execution time with the plane pipelining (Fig.~\ref{fig:bus_latency_compare_diagram}), however, the impact of a higher PIM latency at a larger plane size on the total execution time can be mitigated.
This mitigation becomes prominent when the number of planes being pipelined is large enough.
As shown in Fig.~\ref{fig:htree_and_sharedbus_exeuction_time}b, the increased plane size from \texttt{Size B} to \texttt{Size A} raises the total execution time by 17\% on average with a 2$\times$ higher cell density, i.e., 12.84Gb/mm\textsuperscript{2} for \texttt{Size A}.
Note that we used 64 planes for \texttt{Size A} and 128 planes for \texttt{Size B} to match the PIM throughput (\# of active BLs per cycle) between two cases.

\section{LLM Tiling and Mapping on 3D Flash PIM}\label{sec:op_mapping}

\subsection{Target Operations in LLMs and Their Mapping}\label{sec:llm_op}
\begin{figure}[t]
\centerline{{\includegraphics[width=0.9\columnwidth]{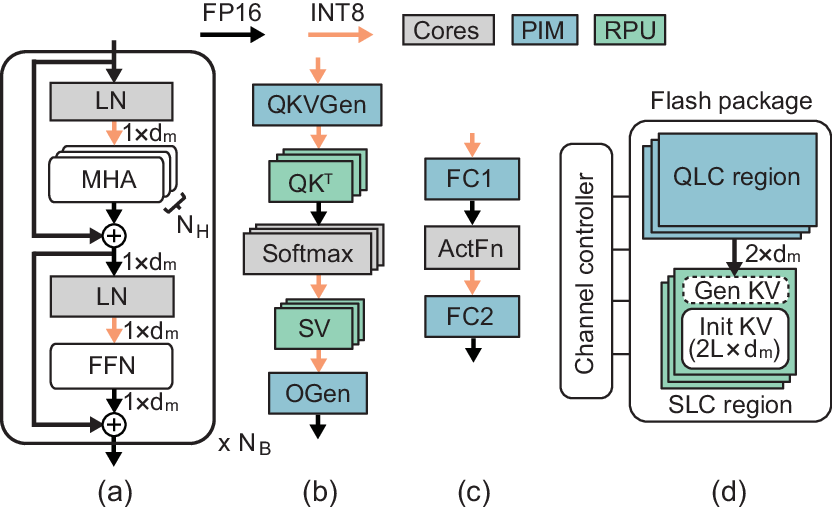}}}
\caption{Overview of LLM layer mapping to compute units in our flash PIM: (a) a decoder block, (b) a multi-head attention (MHA) module (c) a feed-forward network (FFN) and (d) KV caching in QLC-SLC hybrid architecture.}\vspace{-2mm}
\label{fig:llm_layers}
\end{figure}

LLM consists of $N_{\text{B}}$ decoder blocks with $d_{\text{m}}$ hidden dimension, e.g., $N_{\text{B}}\!=\!48$, $d_{\text{m}}\!=\!7,168$ for OPT-30B, in which layer normalization (LN), multi-head attention (MHA), and feed-forward network (FFN) are connected in series (Fig.~\ref{fig:llm_layers}a).
The MHA consists of $N_{\text{H}}$ heads which process an input vector $\mathbf{x}\in\mathbb{R}^{d_\text{m}}$ in parallel.
Fig.~\ref{fig:llm_layers} shows which layers are mapped to which compute units in our 3D flash PIM.
In the PIM array, we adopt the W8A8 quantization scheme, e.g., SmoothQuant~\cite{smoothquant}.
The RPUs handle $QK^\text{T}$ and $SV$ in \texttt{INT16}, while the cores in the SSD controller execute the softmax and activation function in \texttt{FP16}.
The LN layer is also handled in SSD cores as it requires to collect $d_{\text{m}}$ input elements, which are generated across multiple channels for high parallelism.

The layers, except those executed in the cores, are classified into two types of MVM: \textit{static-MVM} (sMVM) and \textit{dynamic-MVM} (dMVM).
The sMVM is the multiplication between weights stored in flash cells (\textit{static}) and an input vector.
The dMVM is the multiplication between \textit{dynamically} generated query ($Q$), key ($K$), and value ($V$).
To effectively support both sMVM and dMVM, we propose to partition dies within a package to a PIM-enabled QLC and a non-PIM SLC region (Fig.~\ref{fig:llm_layers}d).
We map sMVM operations to 3D PIM arrays in the QLC region since no write operations are involved.
Then, the generated $\mathbf{q}$, $\mathbf{k}$, and $\mathbf{v}$ vectors, e.g., $\mathbf{q}=W_Q\!\cdot\!\mathbf{x}$, for the previously generated token are moved to the SLC region for the following $QK^\text{T}$ and $SV$ computations.
The dMVMs are done in RPUs of the SLC region because the programming latency of SLC NAND is $19\times$ lower than that of QLC NAND~\cite{qlc_programming}.
As shown in Fig.~\ref{fig:llm_layers}d, the initial KV is cached in the SLC region while newly generated $\mathbf{k}$ and $\mathbf{v}$ are appended to form a new $K$ and $V$ for next token generation.


\subsection{Tiling and Dataflow of MVMs in Flash}\label{sec:mvm_mapping}
\begin{figure}[t]
\centerline{{\includegraphics[width=0.75\columnwidth]{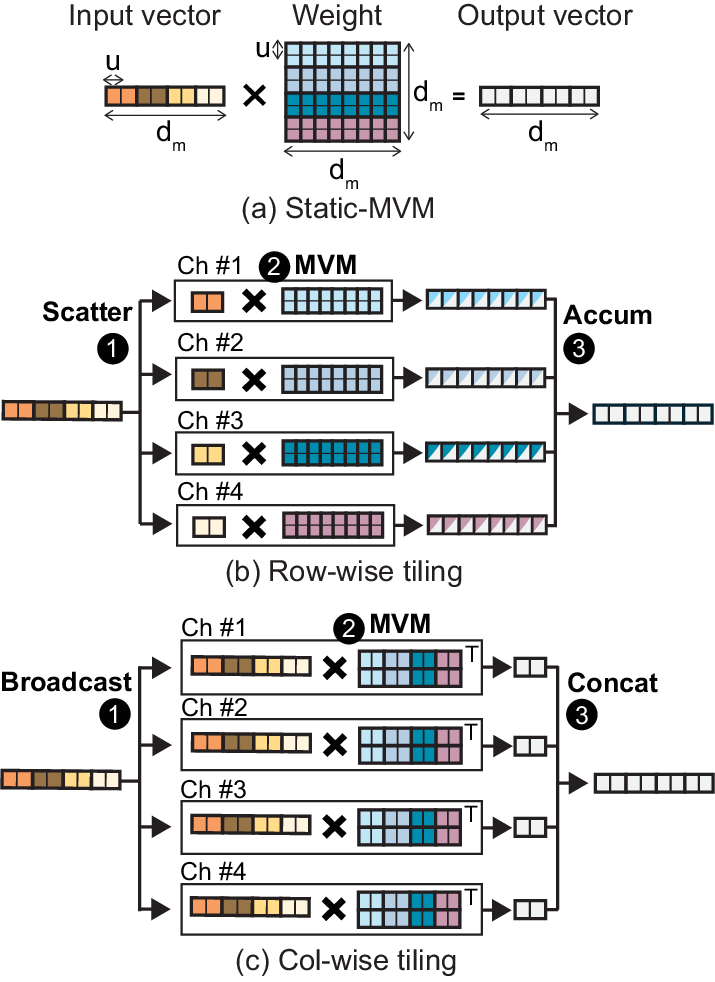}}}
\caption{Example of two different channel-level tiling methods for the static-MVM.}\vspace{-2mm}
\label{fig:staticMVM}
\end{figure}

\textbf{Tiling and dataflow in sMVM:} The NAND flash has four levels of hierarchy, i.e., channel, way, die, and plane, as illustrated in Fig.~\ref{fig:nand_hierarchy}.
Accordingly, PIM parallelism can be realized across four levels.
Considering two ways in tiling weights (row-wise in Fig.~\ref{fig:staticMVM}b and column-wise in Fig.~\ref{fig:staticMVM}c), the number of combinations for tiling weights across the flash hierarchy becomes $2^4=16$.
In addition, the number of resources to be used at each level ranges from 1 to its own count.
For example, PIM could be enabled for four channels among eight for the sMVM operation, as shown in Fig.~\ref{fig:staticMVM}.
In short, when tiling weights for sMVM, two parameters should be determined at every hierarchy level: the tiling method and the resource count (equivalently, a mapped tile count).

As in Fig.~\ref{fig:staticMVM}, the types of operations involved in sMVM depend on the tiling method.
The row-wise tiling scatters an input vector for an MVM and then accumulates MVM results (Fig.~\ref{fig:staticMVM}b).
To scatter the input vector to PIM arrays, the product of tile counts over the four levels should be $d_{\text{m}}/u$, where $u=128$ is the number of weight rows in MVM.
The unit tile size is $u\times\!(N_\text{col}/4)$, which is determined by the plane size for PIM.
Instead, the col-wise tiling broadcasts the input vector and concatenates MVM results (Fig.~\ref{fig:staticMVM}c).
Given that the size of the concatenated MVM result is $d_\text{m}$, the product of tile counts over the four levels should be $d_{\text{m}}/(N_{\text{col}}/4)$.

\begin{figure}[t]
\centerline{{\includegraphics[width=0.8\columnwidth]{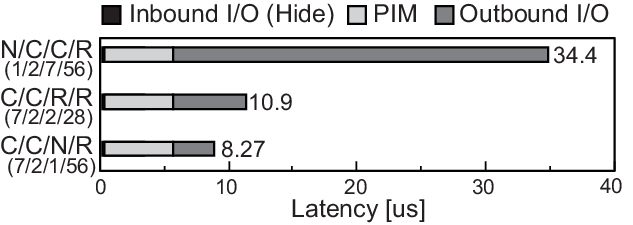}}}
\caption{Latency breakdowns of three tiling options for the static-MVM (N: none, C: column-wise, R: row-wise).}\vspace{-2mm}
\label{fig:breakdown_static_mvm}
\end{figure}

\begin{figure*}[t]
\centerline{{\includegraphics[width=0.9\linewidth]{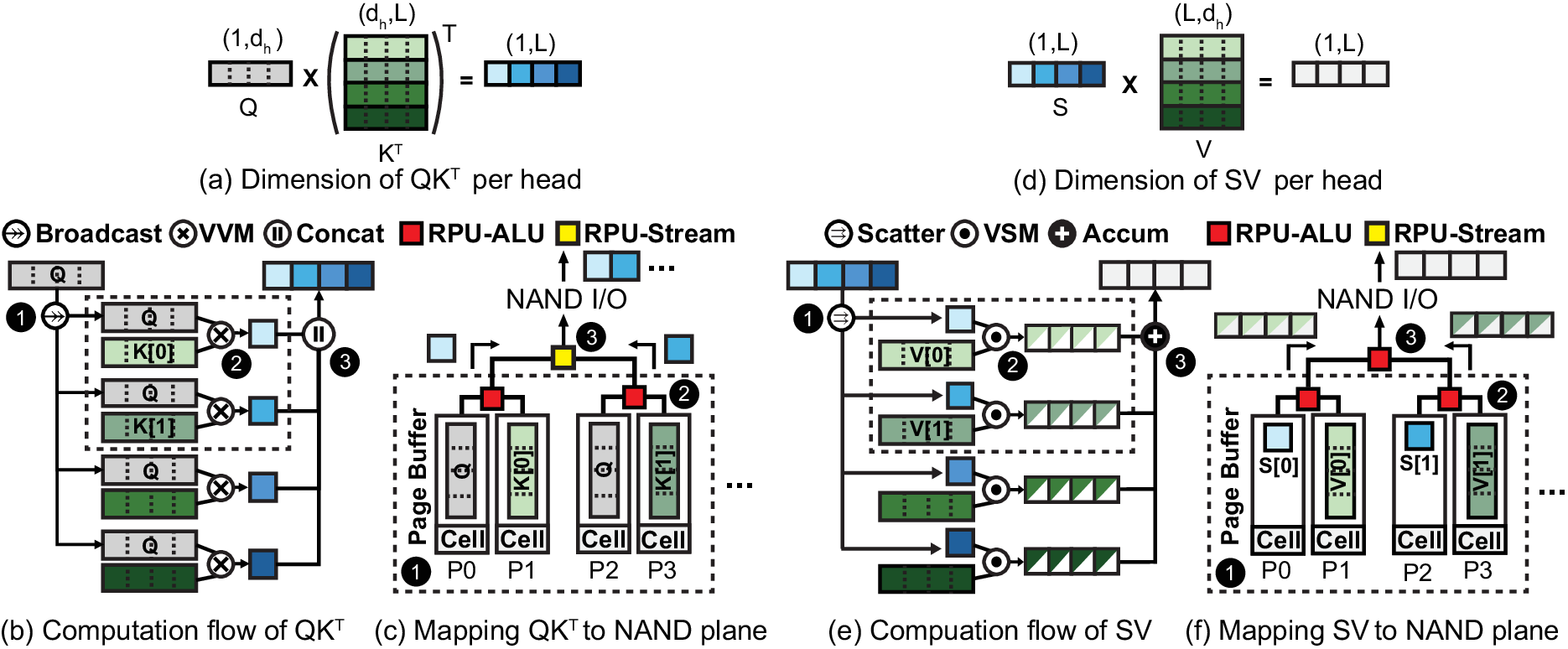}}}
\caption{(a-c) Dimension, computation flow, and mapping of $QK^\text{T}$ per head, and (d-f) dimension, computation flow and mapping of $SV$ per head for single-batch token generation (i.e., $\mathbf{q}\in \mathbb{R}^{1\times d_h}$).}\vspace{-3mm}
\label{fig:qkt_sv_mapping}
\end{figure*}

Fig.~\ref{fig:breakdown_static_mvm} shows the latency breakdowns of the three best cases with $d_m=7,168$ (i.e., OPT-30B) with 8 channels, 4 ways, 8 dies, and 256 planes with \texttt{Size A}.
Each case is denoted by `tiling methods (tile counts)' for every hierarchy level, i.e., \textit{ch}/\textit{way}/\textit{die}/\textit{plane}.
The `N' means no tiling is applied, making the tile count assigned at that level equal to 1.
Since the tile count exploiting the row-wise tiling is equal in all cases, i.e., $56$, both inbound I/O and PIM latencies are identical.
However, using column-wise tiling at the channel level dramatically reduces the outbound I/O latency (`N/C/C/R' vs. the other two).
This is because the number of output tiles that need to move out per channel is reduced.
By exploiting the H-tree network within a die, we may reduce the outbound I/O latency by 47\% when comparing `C/C/R/R' vs. `C/C/N/R').

\textbf{Dataflow of dMVM:} For the token generation, we need to keep writing newly generated $\mathbf{k}$ and $\mathbf{v}$ vectors along with the initial KV cache.
However, data writes can be detrimental to the endurance and performance of the flash cell.
Fortunately, the P/E cycle increases up to 50$\times$ when the cell retention time is reduced to 3 days~\cite{endurance_ssd}.
Considering that the SLC P/E cycle is about $10\text{K}$~\cite{qlc_programming} and TPOT of OPT-30B is about 7ms (Fig.~\ref{fig:conventional_size_nand_pim}), 32GiB SLC can support up to 32 years of LLM running, which is longer than the 5-year warranty of typical SSDs.
This lifetime projection is estimated based on a similar approach in~\cite{optimstore}.

Since the proposed 3D NAND flash PIM targets token generation in LLMs, we need to analyze the overhead of writing the initial KV cache computed by GPUs to a flash device.
With every channel connected to the SLC region, we can utilize `\# of channels$\times$bus speed' for KV cache write.
Given that the sequential write bandwidth in commercial SLC NAND is $4.8-6$GB/s~\cite{slc_micron}, the initial KV cache write for W8A8 OPT-30B with 1K input tokens can be completed in 120ms.
Since generating a single token with OPT-30B on four RTX4090 with vLLM takes 10ms longer than our flash PIM solution (Fig.~\ref{fig:conventional_size_nand_pim}), we can offset the initial KV cache write overhead when generating more than 12 tokens.

To exploit head-level parallelism in MHA, we assign one die per one or two heads of $QK^\text{T}$ and $SV$.
The number of heads assigned depends on the model size.
Obviously, as shown in Fig.~\ref{fig:qkt_sv_mapping}a, $QK^\text{T}$ can be translated into multiple vector-vector multiplications (VVMs).
This is done by broadcasting $\mathbf{q}$ to all rows of matrix $K$, which remains non-transposed in the page buffer (Fig.~\ref{fig:qkt_sv_mapping}b-c).
However, translating $SV$ into the VVM is non-trivial because $L$ increases during the generation stage.
It can be resolved by adopting a row-wise product.
Each element of the S-vector is scattered across planes to perform a vector-scalar multiplication (VSM) with each row of the matrix $V$ (Fig.~\ref{fig:qkt_sv_mapping}e).
Since two operands of each VVM or VSM are stored in a pair of planes, each loading one operand to its page buffer, these multiplications are done in parallel through RPUs in the H-tree network (Fig.~\ref{fig:qkt_sv_mapping}c, Fig.~\ref{fig:qkt_sv_mapping}f).



\section{Experimental Results}\label{sec:exp_results}
\subsection{Simulation Environment}\label{sec:sim_env}
We modeled our flash PIM by combining 3D-FPIM~\cite{3d-fpim} and NeuroSim~\cite{neurosim} simulators to extract PIM latency and power consumption.
Then, we evaluated system-level LLM performance using SimpleSSD simulator~\cite{simplessd} with an in-house simulator to search for the best tiling method (Fig.~\ref{fig:breakdown_static_mvm}).
Specifically, we modified the SimpleSSD simulator to evaluate LN and softmax latencies on ARM cores and used the in-house simulator for sMVM and dMVM latency estimation.
Table~\ref{tbl:simul_params} summarizes the parameters used in these simulations.
Pipeline execution of sMVM comprises three stages: inbound I/O, PIM, and outbound I/O, where the first two overlap.
For dMVM, the same three-stage pipeline is used, but the PIM stage is replaced with KV cache read.
Outbound I/O via H-tree in both sMVM and dMVM involves pipelined execution between RPU and output data transfer.
To hide the accumulation latency in RPUs, we set the clock frequency of RPUs to 250MHz, considering the bus bandwidth.
\subsection{Performance of Proposed 3D Flash PIM on LLM}\label{sec:llm_perf}
\begin{figure}[t]
\centerline{{\includegraphics[width=\columnwidth]{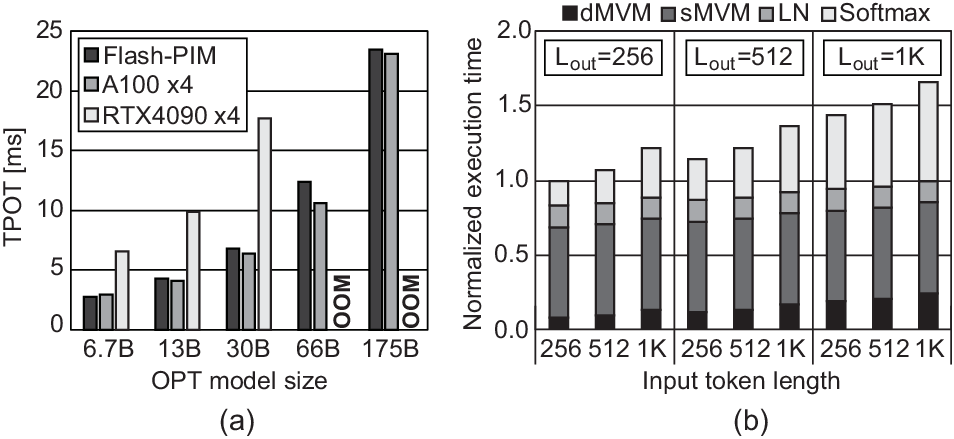}}}
\caption{(a) TPOT comparison across various OPT model sizes and (b) execution time breakdown of our flash PIM by input and output token lengths.}\vspace{-2mm}
\label{fig:flash_vs_gpu}
\end{figure}

To compare LLM single-token generation performance of 3D flash PIM with GPU, four high-end GPUs, i.e., RTX4090, A100, are selected.
As LLM benchmarks, we selected an OPT-family from the smallest one to the largest one, from OPT-6.7B to OPT-175B.
Since the actual deployment of four RTX4090 using vLLM, has not enough VRAM capacity to support OPT-66B and 175B in W8A8 (\textbf{OOM} in Fig.~\ref{fig:flash_vs_gpu}a), AttAcc simulator~\cite{attacc} is used to run larger models with the A100 setup.

The 3D flash PIM achieves lower latency than RTX4090$\times4$ with vLLM and comparable performance with A100$\times4$ (AttAcc) in every OPT model (Fig.~\ref{fig:flash_vs_gpu}a). 
In this evaluation, both input and output token lengths are set to 1K.
The average latency overhead of flash PIM compared to A100$\times4$ (AttAcc) is 4.9\%.
Note that we are using a single flash device for token generation instead of expensive GPU cards.
By offloading the generation task to 3D flash PIM, GPUs can be released for other summarization requests, saving the cost.
When processing $QK^\text{T}$ and $SV$, i.e., dMVM in Fig.~\ref{fig:flash_vs_gpu}b, flash PIM is well scalable to increased input/output token length owing to the head-level parallelism and the row-wise product dataflow (Fig.~\ref{fig:qkt_sv_mapping}).
Execution time of other operations except softmax, i.e., sMVM and LN in Fig.~\ref{fig:flash_vs_gpu}b, is consistent regardless of input/output token length because it depends on the model dimension, not token lengths.

\subsection{Area Analysis of Our 3D Flash PIM}\label{sec:area_overhead}
Considering BGA316 (14mm$\times$18mm) allows up to 32 dies and these dies are stacked within the package, the budget area of the die can be estimated.
When four dies are stacked with a 60\% overlap and the dies occupy 30–40\% of the BGA316, the estimated budget area per die ranges $5.6\,–\,7.5\text{mm}^2$.
This is larger than the total area of 256 flash PIM arrays in the selected size of $256\!\times\!2048\!\times\!128$, which is 4.98$\text{mm}^2$.
Moreover, recent 3D NAND architectures adopt the peri-under-array (PUA) structure to improve cell density~\cite{isscc_19_pua}, in which peripheral circuits are located under NAND plane's memory array.
According to~\cite{pua_pba}, the low-voltage peripheral circuits (LV-peri) can be scaled down to an advanced technology node, e.g., 7nm, reducing its area to 23.16\% of the plane (Table~\ref{tbl:area_breakdown}).
We synthesized RPUs using Synopsys Design Compiler with 65nm and scaled its area down to 7nm.
The area analysis of H-tree network is estimated using 7nm metal-1 (M1) pitch size and wiring length required to connect 256 planes.
Both reported values are normalized by the number of planes in a die, i.e., 256 planes.
The flash PIM peripheral circuits and the H-tree network with RPUs account for less than 50\% of the plane size,
which can be successfully integrated under the memory array without incurring additional area overhead.


\begin{table}[t]
\caption{Simulation parameters}
\vspace{-2mm}
\label{tbl:simul_params}
\renewcommand{\arraystretch}{1.1}
\resizebox{\columnwidth}{!}{%
\begin{tabular}{|l|l|}
\hline
\multicolumn{1}{|c|}{Components} & \multicolumn{1}{c|}{Configuration}                                                                                     \\ \hline \hline
Controller                       & 4 ARM Cortex-A9 cores, PCIe 5.0 $\times$4                                                                                     \\ \hline
Form factor                      & NGSFF(Next Generation Small Form Factor), BGA316 x8                                                                                                       \\ \hline
\begin{tabular}[c]{@{}l@{}}Flash \\ memory\end{tabular} &
  \begin{tabular}[c]{@{}l@{}}8 channels, 4 ways, 8 dies (2:SLC, 6:QLC), 256 planes \\ Page size$=$256B, 4 BLSs per block, 64 blocks, 128 stacks\\ flash bus$=$2GB/s (1000MT/s, 8-bit)\end{tabular} \\ \hline
RPU                              & \begin{tabular}[c]{@{}l@{}}250MHz, INT16 multiplier $\times$8, INT32 adder $\times$9\\ 64-bit register $\times$5, 256-bit register $\times$1,\end{tabular} \\ \hline
\end{tabular}\vspace{-2mm}
}
\end{table}

\begin{table}[t]
\centering
\caption{Area breakdown of peripheral circuits and H-tree network with RPUs per plane}\vspace{-2mm}
\label{tbl:area_breakdown}
\renewcommand{\arraystretch}{1.1}
\footnotesize 
\resizebox{0.95\columnwidth}{!}{%
\begin{tabular}{|l|l|l|l|}
\hline
                    & HV-peri + cap   & LV-peri   & RPU + H-tree    \\ \hline\hline
Area {[}$mm^2${]}    & 0.004210  & 0.004510  & 0.000077 \\ \hline
Area ratio in plane & 21.62\%   & 23.16\%   & 0.39\%    \\ \hline
\end{tabular}
}
\\
\vspace{1mm} 
\raggedright
\scriptsize
\hspace*{3mm}LV-peri = {decBLS, precharger, mux, ADC, page buffer, shiftadder} \\
\vspace{0.5mm} 
\hspace*{3mm}HV-peri = decWL
\end{table}


\section{Conclusion}
This work presents a novel 3D NAND flash PIM architecture optimized for single-batch token generation in LLMs.
The optimal plane size of $256\!\times\!2048\!\times\!128$ for 3D flash PIM is determined through extensive design space exploration for low PIM latency.
In addition, H-tree network is being utilized to support on-the-fly output accumulation for both sMVM and dMVM.
Proper operation tiling and novel mapping of dMVM and sMVM on the proposed QLC-SLC hybrid architecture enables 2.4$\times$ speedup over four RTX4090 and comparable performance to four A100 with only 4.9\% latency overhead on average.
Notably, this low-cost and fast flash PIM solution is feasible without incurring any area overhead.

\section*{ACKNOWLEDGMENTS}
This work was partially supported by the Institute of Information \& Communications Technology Planning \& Evaluation (IITP) funded by the Ministry of Science and ICT under Grant 2022-0-01170 and Grant RS-2023-00229849; in part by the National Research Foundation of Korea (NRF) funded by the Ministry of Science and ICT under Grant RS-2023-00258227.

\bibliographystyle{IEEEtran}
\bibliography{refs}

@inproceedings{transformer,
 author = {Vaswani, Ashish and others},
 booktitle = {NeurIPS},
 pages = {},
 title = {Attention is All you Need},
 year = {2017},
}

@inproceedings{gpt3,
 author = {Brown, Tom and others},
 booktitle = {NeurIPS},
 title = {Language Models are Few-Shot Learners},
 year = {2020},
}

@article{mixtral,
    title	= {Mixtral of Experts},
    author	= {Albert Q. Jiang and others},
    year	= {2024},
    journal	= {arXiv:2401.04088}
}

@INPROCEEDINGS{nand_isscc25,
  author={Cho, Wanik and others},
  booktitle={ISSCC}, 
  title={A 321-Layer {2Tb} 4b/cell {3D-NAND-Flash} Memory with a {75MB/s} Program Throughput}, 
  year={2025},
}

@INPROCEEDINGS{flash_pim_ucsb,
  author={Bavandpour, Mohammad and others},
  booktitle={DATE}, 
  title={Mixed-Signal Vector-by-Matrix Multiplier Circuits Based on {3D-NAND} Memories for Neurocomputing}, 
  year={2020},
}

@ARTICLE{jssc_flash_pim,
  author={Kim, Minsu and others},
  journal={IEEE JSSC}, 
  title={An Embedded {NAND} Flash-Based Compute-In-Memory Array Demonstrated in a Standard Logic Process}, 
  year={2022},
}

@INPROCEEDINGS{3d-fpim,
  author={Lee, Hunjun and others},
  booktitle={MICRO}, 
  title={{3D-FPIM}: An Extreme Energy-Efficient {DNN} Acceleration System Using {3D NAND} Flash-Based In-Situ {PIM} Unit}, 
  year={2022},
  volume={},
  number={},
}

@ARTICLE{pua_pba,
  author={Shim, Wonbo and others},
  journal={IEEE EDL}, 
  title={Technological Design of {3D NAND}-Based Compute-in-Memory Architecture for {GB}-Scale Deep Neural Network}, 
  year={2021},
}

@inproceedings{attacc,
author = {Park, Jaehyun and others},
title = {{AttAcc}! Unleashing the Power of {PIM} for Batched Transformer-based Generative Model Inference},
booktitle = {ASPLOS},
year = {2024},
}

@INPROCEEDINGS{zssd,
  author={Cheong, Wooseong and others},
  booktitle={ISSCC}, 
  title={A flash memory controller for 15us ultra-low-latency {SSD} using high-speed {3D NAND} flash with 3us read time}, 
  year={2018},
  volume={},
  number={},
  doi={10.1109/ISSCC.2018.8310322}
}

@inproceedings{vllm,
author = {Kwon, Woosuk and others},
title = {Efficient Memory Management for Large Language Model Serving with {PagedAttention}},
booktitle = {SOSP},
year = {2023},
}

@INPROCEEDINGS{isscc_18,
  author={Maejima, Hiroshi and others},
  booktitle={ISSCC}, 
  title={A 512{Gb} 3b/Cell {3D} flash memory on a 96-word-line-layer technology}, 
  year={2018},
}

@ARTICLE{neurosim,
  author={Chen, Pai-Yu and others},
  journal={IEEE TCAS}, 
  title={NeuroSim: A Circuit-Level Macro Model for Benchmarking Neuro-Inspired Architectures in Online Learning}, 
  year={2018},
}

@book{book_3dflash,
author = {Micheloni, Rino},
title = {3D Flash Memories},
year = {2016},
isbn = {9401775109},
publisher = {Springer},
edition = {1st},
}

@inproceedings{networkedSSD,
author = {Kim, Jiho and others},
booktitle={MICRO}, 
title={{Networked SSD}: Flash Memory Interconnection Network for High-Bandwidth {SSD}}, 
year={2022},
}

@misc{horowitz,
  title={Timing Models for {MOS} Circuits},
  author={Mark Horowitz},
  year={1983},
  url={https://api.semanticscholar.org/CorpusID:60771551},
}

@inproceedings{smoothquant,
author = {Xiao, Guangxuan and others},
title = {{SmoothQuant}: accurate and efficient post-training quantization for large language models},
year = {2023},
booktitle = {ICML},
articleno = {1585},
numpages = {13},
}

@INPROCEEDINGS{qlc_programming,
  author={Takai, Yoshiki and others},
  booktitle={IEEE IMW}, 
  title={Analysis on Heterogeneous {SSD} Configuration with Quadruple-Level Cell ({QLC}) {NAND} Flash Memory}, 
  year={2019},

  doi={10.1109/IMW.2019.8739689}
}

@INPROCEEDINGS{optimstore,
  author={Kim, Junkyum and others},
  booktitle={HPCA}, 
  title={{OptimStore}: In-Storage Optimization of Large Scale {DNNs} with On-Die Processing}, 
  year={2023},
  volume={},
  number={},
  pages={611-623},
  doi={10.1109/HPCA56546.2023.10071024}}

@INPROCEEDINGS{endurance_ssd,
  author={Luo, Yixin and others},
  booktitle={MSST}, 
  title={{WARM}: Improving {NAND} flash memory lifetime with write-hotness aware retention management}, 
  year={2015},
}

@online{slc_micron,
  title        = {Micron {XTR} {NVMe}\texttrademark{} {SSD} — Product Brief},
  author       = {{Micron Technology, Inc.}},
  year         = {2023},
  url          = {https://in.micron.com/content/dam/micron/global/public/products/product-flyer/xtr-nvme-ssd-product-brief.pdf},
  urldate      = {2025-05-24}
}

@INPROCEEDINGS{simplessd,
  author={Gouk, Donghyun and others},
  booktitle={MICRO}, 
  title={Amber: Enabling Precise Full-System Simulation with Detailed Modeling of All {SSD} Resources}, 
  year={2018},
}

@INPROCEEDINGS{isscc_19_pua,
  author={Siau, Chang and others},
  booktitle={ISSCC}, 
  title={A 512{Gb} 3-bit/Cell {3D} Flash Memory on 128-Wordline-Layer with 132MB/s Write Performance Featuring Circuit-Under-Array Technology}, 
  year={2019},
}

\end{document}